\documentstyle[epsfig,amssymb,aps]{revtex}
\begin{document}
\newcommand{\mafigura}[4]{
  \begin{figure}[hbtp]
    \begin{center}
      \epsfxsize=#1 \leavevmode \epsffile{#2}
    \end{center}
    \caption{#3}
    \label{#4}
  \end{figure} }
\author{L. B. Leinson$^{1}$ and A. P\'{e}rez}
\address{Departamento de F\'{\i}sica Te\'{o}rica, Universidad de Valencia \\
46100 Burjassot (Valencia), Spain\\
$^{1}$On leave from : Institute of Terrestrial Magnetism, Ionosphere and\\
Radio Wave Propagation\\
RAS, 142092 Troitsk, Moscow Region, Russia }
\title{Neutrino-pair radiation form neutron star crusts: Collective effects}
\maketitle

\begin{abstract}
We present a method to calculate $\nu \bar{\nu}$ energy losses from neutron
star crusts, which automatically takes into account for collective effects,
and allows to calculate the total emissivity without a separate
consideration of particular processes. We show that the formula we obtain
describe the known results for the emissivity due to plasmon decay and
Bremsstrahlung from degenerate electrons, when one of this processes
dominates. In the case of low temperatures, our formula gives a suppression
of the electron vector weak-current contribution to $\nu \bar{\nu}$
Bremsstrahlung, due to the collective effects discussed in this paper.
\end{abstract}


\section{Introduction}

\noindent According to modern theoretical scenarios, within the first
hundred years of existence of a neutron star the cooling of its interior
layers occurs in an extremely nonuniform way. Due to the so-called ''fast
cooling processes'' in nuclear matter, the core of a neutron star cools down
very fast, while the external crust cools down much slowly, by heat
diffusion towards the inside and by neutrino radiation \cite{Latt94}. The
time necessary for thermal relaxation crucially depends on the intensity of
neutrino and antineutrino radiation from the crust of the neutron star,
which influences the temperature of the stellar surface, and thus the gamma
and X-ray which are observed. It is well-known that radiation of neutrinos
and antineutrinos from the neutron star crust is caused dominantly by $\nu 
\bar{\nu}$ Bremsstrahlung of degenerate electrons, and by plasmon decay into
neutrino pairs. The Bremsstrahlung process, operating in a liquid or
crystalline phase of the crust, has been previously studied by many authors 
\cite{FR69} - \cite{YK96}. These calculations have been performed neglecting
collective interactions of electrons with the neutrino field. When the
temperature of the crust is smaller than the electron plasma frequency, such
an approximation is not justified because, for the scenario under
consideration, the wavelength $\lambda $ of radiated neutrinos and
antineutrinos is larger than the electron Debye screening distance $D_{e}$.
By undergoing a quantum transition, the radiating electron
electromagnetically induces a motion of other electrons inside the Debye
sphere around itself. The weak current of perturbed electrons inside the
Debye sphere generates neutrinos coherently with the initial electron,
therefore screening its vector weak coupling with the neutrino field. Since
this collective effect takes place when the neutrino wave-length is of the
order, or larger, than the Debye screening distance, the relevant parameter
of the problem is $k^{2}D_{e}^{2}$, where $k$ is the momentum carried out by
the neutrino-pair\footnote{%
In what follows we use the system of units $\hbar =c=1$ and the Boltzmann
constant $k_{B}=1$. The fine-structure constant is $\alpha =e^{2}=1/137$.}.
The screening effect was demonstrated in \cite{L99} for neutrino-pair
emission due to electron-phonon scattering in\ a crystalline crust, where
the condition $k^{2}D_{e}^{2}\ll 1$ is fulfilled. In this limiting case,
valid when the temperature $T$\ is much smaller than the electron plasma
frequency $\omega _{pe}$, the effective vector weak-current of electrons is
totally screened by the plasma polarization, which dramatically modifies the
neutrino emissivity.

In the present paper, we introduce a method of calculation for $\nu \bar{\nu}
$ energy losses from a neutron star crust, which incorporates collective
effects in a degenerate plasma at arbitrary temperatures, limited only by
the degeneration condition $T\ll \mu _{e}$ , where $\mu _{e}$ is the
electron chemical potential. This condition holds during the cooling epoch
described above.

Collective effects appear if one takes into account the possibility of
virtual photon-exchange among electrons in the plasma. To obtain the
neutrino emissivity, we use the fluctuation-dissipation theorem in order to
relate the weak current-current correlation function to the imaginary part
of the exact retarded polarization functions of the plasma. This procedure,
as we will see, has the advantage that collective effects are automatically
included, and do not need a separate consideration. The obtained formula for
the neutrino energy losses includes contribution of plasmon decay into
neutrino pairs, as well as the process of neutrino-pair emission due to
electron collisions with nuclei and phonons. We identify the latter
mechanism as the\ Bremsstrahlung process from electrons. Thus, the total
neutrino-pair emissivity of the crust, caused by $\nu \bar{\nu}$ decay of
plasmons and Bremsstrahlung from electrons is represented by a unique
expression.

This paper is organized as follows. In Sect. 2 we discuss the reference
status of the problem, and derive some formulae describing the rate of $\nu 
\bar{\nu}$ Bremsstrahlung, which we use in the following considerations. In
Sect. 3, we show how the same $\nu \bar{\nu}$ Bremsstrahlung rate can be
obtained by the use of the Optical Theorem, and calculate the imaginary
parts of the retarded polarization tensors of a liquid or crystallin crust.
In order to incorporate collective effects, in Sec.4 we generalize the
results\ obtained by the Optical Theorem by replacing the imaginary parts of
the polarization tensors by the weak current-current correlation function,
according to the Fluctuation-Dissipation Theorem. This correlation function
is obtained by summation of all diagrams with allowance of intermediate
photon-exchange among the electrons. In Sect. 5 we derive a general formula
for the $\nu \bar{\nu}$\ emissivity, which includes the energy losses due to
electron Bremsstrahlung and plasmon decay. Some limiting cases and numerical
tests are presented in Sect. 6 in order to demonstrate the validity of the
obtained general formula. Discussion of the results and conclusions are
shown in Sect. 7.

\section{Statement of the problem}

We use the Standard Model of weak interactions, and consider low-energy
electrons, which are typical for neutron star interiors. Therefore the
in-vacuum weak interaction of electrons with the neutrino field can be
written in a point-like current-current approach 
\begin{equation}
{\cal H}_{eff}=\frac{G_{F}}{\sqrt{2}}\int j^{\mu }\left( x\right) J_{\mu
}\left( x\right) d^{4}x,  \label{H}
\end{equation}
where $G_{F}$ is the Fermi coupling constant, and 
\[
j^{\mu }=\bar{\nu}\text{\thinspace }\gamma ^{\mu }\left( 1-\gamma
_{5}\right) \,\nu 
\]
is the neutrino current. The vacuum weak current of an electron is of the
standard form, which is a sum of vector and axial-vector pieces

\begin{equation}
J_{\mu }=\bar{\psi}\left( C_{V}\gamma _{\mu }-C_{A}\gamma _{\mu }\gamma
_{5}\right) \psi .  \label{J}
\end{equation}
Here, $\psi $ stands for electron field; $C_{V}=\frac{1}{2}+2\sin ^{2}\theta
_{W}$ , $C_{A}=\frac{1}{2}$ stand for emission of electron neutrinos,
whereas $C_{V}^{\prime }=-\frac{1}{2}+2\sin ^{2}\theta _{W}$ , $%
C_{A}^{\prime }=-\frac{1}{2}$\ are to be used for muon and tau neutrinos; $%
\theta _{W}$ is the Weinberg angle.

The differential rate of $\nu \bar{\nu}$ production takes the form 
\begin{eqnarray}
d\Gamma &=&\frac{G_{F}^{2}}{2}\;\frac{1}{\left( 2\pi \right) ^{6}}\int \frac{%
d^{3}p}{2E}\int \frac{d^{3}p^{\prime }}{2E^{\prime }}f\left( E\right) \left[
1-f\left( E^{\prime }\right) \right]  \nonumber \\
&&%
\mathop{\rm Tr}%
\left( M_{\mu }^{\dagger }M_{\nu }\right) 
\mathop{\rm Tr}%
\left( j^{\mu }j^{\nu \ast }\right) \frac{d^{3}k_{1}}{2\omega _{1}(2\pi )^{3}%
}\frac{d^{3}k_{2}}{2\omega _{2}(2\pi )^{3}}.  \label{rate2}
\end{eqnarray}
Here $M_{\mu }$ is the matrix element of the weak transition current for the
electron in the Bremsstrahlung process. The symbol $%
\mathop{\rm Tr}%
\left( M_{\mu }^{\dagger }M_{\nu }\right) $ includes statistical averaging
and summation over initial and final states of the background (see below).
Summation over all initial $p=\left( E,{\bf p}\right) $ and final $p^{\prime
}=\left( E^{\prime },{\bf p}^{\prime }\right) $ states of the electron has
to take into account the Pauli principle via the appropriate blocking
factors, with the Fermi distribution function $f\left( E\right) $ of
degenerate electrons. Finally, $k_{1}=(\omega _{1},{\bf k}_{1})$ , $%
k_{2}=(\omega _{2},{\bf k}_{2})$ are the neutrino and antineutrino
four-momenta, respectively. To simplify the calculations in what follows, we
consider an ultrarelativistic electron gas, which is typical for the
dominant volume of the neutron star crust.

The matrix element for the weak transition current of the electron has the
following form \cite{Fl73}: 
\begin{eqnarray}
M^{\mu } &=&\bar{u}\left( p^{\prime }\right) \left[ \gamma ^{\mu }\left(
C_{V}-C_{A}\gamma _{5}\right) G\left( p^{\prime }+K\right) \gamma ^{\alpha
}+\gamma ^{\alpha }G\left( p-K\right) \gamma ^{\mu }\left( C_{V}-C_{A}\gamma
_{5}\right) \right] u\left( p\right)  \nonumber \\
&&\times \int_{-\infty }^{\infty }\,dte^{-iq_{0}t}\left\langle f\right|
A_{\alpha }\left( -{\bf q,t}\right) b^{\dagger }\left( p^{\prime },\sigma
^{\prime }\right) b\left( p,\sigma \right) c^{\dagger }\left( k_{1}\right)
d^{\dagger }\left( k_{2}\right) \left| i\right\rangle .
\end{eqnarray}
In this equation, $p=\left( E,{\bf p}\right) $ and $p^{\prime }=\left(
E^{\prime },{\bf p}^{\prime }\right) $ are the initial and final momenta of
the electron. The operators b, c, d refer to electrons, neutrinos and
antineutrinos, respectively. In this process, the momentum and energy
transferred to the background are ${\bf q=p-p}^{\prime }-{\bf k}$ and $%
q_{0}=E-E^{\prime }-\omega ,$ where $K=\left( \omega ,{\bf k}\right) $ is
the total four-momentum of the neutrino pair : $K=k_{1}+k_{2}$. . Within the
Coulomb gauge, used in this section, the electromagnetic potential has only
a non-vanishing (scalar) component.

The energy transfer to the background, $q_{0}$, is small with respect to the
momentum transfer $\left| {\bf q}\right| $, and is therefore neglected in
the electron matrix element. On the other hand, the momentum transfer to the
background, of the order $\left| {\bf q}\right| \sim p_{F}$\ , where $p_{F}$%
\ is the electron Fermi momentum, is large with respect to the neutrino-pair
momentum, therefore we assume that $q=p-p^{\prime }-k\simeq p-p^{\prime }$.
\ Keeping this in mind, we can use the well-known method of soft photons 
\cite{BLP82}. With the aid of commutation rules and the Dirac equation, one
can represent the matrix element as follows 
\begin{eqnarray}
M^{\mu } &=&\left( \frac{p^{\prime \mu }}{p^{\prime }K}-\frac{p^{\mu }}{pK}%
\right) \bar{u}\left( p^{\prime }\right) \gamma ^{0}\left( C_{V}-C_{A}\gamma
_{5}\right) u\left( p\right) \,  \nonumber \\
&&\times \int_{-\infty }^{\infty }\,dte^{-iq_{0}t}\left\langle f\right|
A_{0}\left( -{\bf q,t}\right) b^{\dagger }\left( p^{\prime },\sigma ^{\prime
}\right) b\left( p,\sigma \right) c^{\dagger }\left( k_{1}\right) d^{\dagger
}\left( k_{2}\right) \left| i\right\rangle .  \label{MatrEl}
\end{eqnarray}
After squaring and taking the trace, and performing summation over final
states and thermal averaging over initial states of the background we obtain 
\begin{eqnarray}
\mathop{\rm Tr}%
\left( M^{\dagger \mu }M^{\nu }\right)  &=&4\left(
C_{V}^{2}+C_{A}^{2}\right) \int dq_{0}d^{3}q\delta \left( {\bf p}-{\bf p}%
^{\prime }-{\bf q-k}\right) \delta \left( E-E^{\prime }-q_{0}-\omega \right) 
\nonumber \\
&&\left( 2EE^{\prime }-pp^{\prime }\right) \left( \frac{p^{\prime \mu }}{%
p^{\prime }K}-\frac{p^{\mu }}{pK}\right) \left( \frac{p^{\prime \nu }}{%
p^{\prime }K}-\frac{p^{\nu }}{pK}\right) \left\langle
A_{0}A_{0}\right\rangle _{q_{0,}{\bf q}},  \label{trace}
\end{eqnarray}
The Fourier transform of the $AA$ correlation function can be written with
the aid of the dynamic form factor $S\left( q_{0},{\bf q}\right) $ : 
\begin{equation}
\left\langle A_{0}A_{0}\right\rangle _{q_{0,}{\bf q}}=\frac{\left( 4\pi
e^{2}Z\right) ^{2}}{\left( q^{2}\varepsilon \left( {\bf q}\right) \right)
^{2}}S\left( q_{0},{\bf q}\right) .
\end{equation}
The dielectric function of a degenerate ultrarelativistic electron gas \cite
{Jan62} can be taken in the static limit 
\begin{equation}
\varepsilon \left( {\bf q}\right) =1+\frac{1}{{\bf q}^{2}D_{e}^{2}}\left( 
\frac{2}{3}+\frac{1-3x^{2}}{6x}\ln \left| \frac{1+x}{1-x}\right| +\frac{x^{2}%
}{3}\ln \left| \frac{x^{2}}{1-x^{2}}\right| \right) ,
\end{equation}
where 
\begin{equation}
x=\frac{\left| {\bf q}\right| }{2p_{F}}.
\end{equation}
Using the approximations discussed above, we can rewrite Eq. (\ref{trace})
as follows 
\begin{eqnarray}
\mathop{\rm Tr}%
\left( M^{\dagger \mu }M^{\nu }\right)  &=&4\left(
C_{V}^{2}+C_{A}^{2}\right) \int dq_{0}d^{3}q\left( \frac{4\pi e^{2}Z}{%
q^{2}\varepsilon \left( q\right) }\right) ^{2}S\left( q_{0},{\bf q}\right)
\left( 2EE^{\prime }-pp^{\prime }\right)   \nonumber \\
&&\times \left( \frac{p^{\prime \mu }}{p^{\prime }K}-\frac{p^{\mu }}{pK}%
\right) \left( \frac{p^{\prime \nu }}{p^{\prime }K}-\frac{p^{\nu }}{pK}%
\right) \delta \left( {\bf p}-{\bf p}^{\prime }-{\bf q}\right) \delta \left(
E-E^{\prime }-q_{0}-\omega \right) .  \label{Msq}
\end{eqnarray}
In a liquid crust, the energy transfer to the background is negligible due
to the large nucleus mass $M_{i}$. In this case we can neglect $q_{0}$
everywhere except in the dynamic form factor. This allows the integration
over $dq_{0}$ to be performed. We obtain 
\begin{eqnarray}
\mathop{\rm Tr}%
\left( M^{\dagger \mu }M^{\nu }\right) _{{\sf liquid}} &=&8\pi N_{i}\left(
C_{V}^{2}+C_{A}^{2}\right) \int d^{3}q\left( \frac{4\pi e^{2}Z}{%
q^{2}\varepsilon \left( q\right) }\right) ^{2}S_{st}\left( {\bf q}\right)
\left( 2EE^{\prime }-pp^{\prime }\right)   \nonumber \\
&&\times \left( \frac{p^{\prime \mu }}{p^{\prime }K}-\frac{p^{\mu }}{pK}%
\right) \left( \frac{p^{\prime \nu }}{p^{\prime }K}-\frac{p^{\nu }}{pK}%
\right) \delta \left( {\bf p}-{\bf p}^{\prime }-{\bf q}\right) \delta \left(
E-E^{\prime }-\omega \right) .  \label{Ltrace}
\end{eqnarray}
The static structure factor of ions is defined as 
\begin{equation}
S_{st}\left( {\bf q}\right) =\frac{1}{N_{i}}\int \frac{dq_{0}}{2\pi }S\left(
q_{0},{\bf q}\right) ,
\end{equation}
where $N_{i}$ is the number density of ions. We use the structure factor
calculated in \cite{Itoh83a} for a one-component classical plasma, which is
a tabulated function of the non-ideality parameter $\Gamma $ of the ionic
component of the plasma.

In the case of a crystallin crust, the dynamic form factor can be written,
in the one-phonon approximation \cite{Fl73} :

\begin{equation}
S\left( q_{0},{\bf q}\right) =2\pi \frac{N_{i}}{M_{i}}e^{-2W\left( q\right)
}\sum_{{\bf s}\lambda {\bf K}}\frac{\left( {\bf q\cdot \hat{e}}_{\lambda 
{\bf s}}\right) }{2\omega _{\lambda {\bf s}}}\left[ \frac{\delta \left(
q_{0}-\omega _{\lambda {\bf s}}\right) \delta _{{\bf q,s-K}}}{1-e^{-\omega
_{\lambda {\bf s}}/T}}+\frac{\delta \left( q_{0}+\omega _{\lambda {\bf s}%
}\right) \delta _{{\bf q,-s-K}}}{e^{-\omega _{\lambda {\bf s}}/T}-1}\right] .
\label{Sph}
\end{equation}
Here $\omega _{\lambda {\bf s}}$ is the phonon frequency, which depends on
the wave vector ${\bf s}$. Each phonon mode $\lambda $ is defined by its
polarization vector ${\bf \hat{e}}_{\lambda {\bf s}}$. The ${\bf K}$ 's are
reciprocal lattice vectors. The Debye-Waller factor is of the form 
\begin{equation}
2W\left( q\right) =\sum_{{\bf s}\lambda {\bf K}}\frac{\left( {\bf q\cdot 
\hat{e}}_{\lambda {\bf s}}\right) ^{2}}{2N_{i}M_{i}\omega _{\lambda {\bf s}}}%
\coth \left( \frac{\omega _{\lambda {\bf s}}}{2T}\right)  \label{DW}
\end{equation}

\section{Treatment by the Optical Theorem}

With the help of the Optical Theorem, the rate Eq. (\ref{rate2}) of $\nu 
\bar{\nu}$ production can be written as 
\begin{equation}
d\Gamma =\frac{G_{F}^{2}}{2}\;\frac{2}{\exp \left( \frac{\omega }{T}\right)
-1}%
\mathop{\rm Im}%
\left[ \Pi _{\mu \nu }\left( K\right) 
\mathop{\rm Tr}%
\left( j^{\mu }j^{\nu \ast }\right) \right] \frac{d^{3}k_{1}}{2\omega
_{1}(2\pi )^{3}}\frac{d^{3}k_{2}}{2\omega _{2}(2\pi )^{3}},  \label{rate1}
\end{equation}
The exact, irreducible retarded polarization tensor $\Pi ^{\mu \nu }$ of the
medium, represents the sum of compact diagrams which include inside the
electromagnetic interactions of the electron with nuclei, and have ends at
the weak vertex $\left( C_{V}\gamma _{\mu }-C_{A}\gamma _{\mu }\gamma
_{5}\right) $. This tensor is the following sum of vector and axial pieces 
\begin{equation}
\Pi ^{\mu \nu }\left( K\right) =\frac{C_{V}^{2}}{4\pi e^{2}}\Pi _{V}^{\mu
\nu }+\frac{C_{A}^{2}}{4\pi e^{2}}\Pi _{A}^{\mu \nu }.  \label{IrrP}
\end{equation}
We traditionally include an extra-factor $4\pi e^{2}$ in the definition of
all polarization tensors. By this reason, the factor $1/4\pi e^{2}$ has been
included in Eq. (\ref{IrrP}). Following these notations, the vector-vector
tensor $\Pi _{V}^{\mu \nu }$ is the retarded tensor for the electromagnetic
polarization of the plasma. In the absence of external magnetic fields, the
parity-violating axial-vector polarization does not contribute to the rate
of neutrino-pair production. In fact, by inserting $\int d^{4}K\delta
^{\left( 4\right) }\left( K-k_{1}-k_{2}\right) =1$ in this equation, and
making use of the Lenard's integral \ 
\begin{eqnarray}
&&\int \frac{d^{3}k_{1}}{2\omega _{1}}\frac{d^{3}k_{2}}{2\omega _{2}}\delta
^{\left( 4\right) }\left( K-k_{1}-k_{2}\right) 
\mathop{\rm Tr}%
\left( j^{\mu }j^{\nu \ast }\right)  \nonumber \\
&=&\frac{4\pi }{3}\left( K_{\mu }K_{\nu }-K^{2}g_{\mu \nu }\right) \theta
\left( K^{2}\right) \theta \left( K^{0}\right) ,
\end{eqnarray}
where $\theta (x)$ is the Heaviside step function, Eq. (\ref{rate1}) can be
written as 
\begin{equation}
d\Gamma =\frac{4\pi }{3}\frac{G_{F}^{2}}{2}\;\frac{2}{\exp \left( \frac{%
\omega }{T}\right) -1}%
\mathop{\rm Im}%
\Pi ^{\mu \nu }\left( K\right) \left( K_{\mu }K_{\nu }-K^{2}g_{\mu \nu
}\right) \theta \left( K^{2}\right) \theta \left( K^{0}\right) \frac{d^{4}K}{%
(2\pi )^{6}}.  \label{rate11}
\end{equation}
Since the axial-vector polarization has to be an antisymmetric tensor, its
contraction in (\ref{rate11}) with the symmetric tensor $K_{\mu }K_{\nu
}-K^{2}g_{\mu \nu }$ vanishes.

\subsection{Polarization tensors}

To specify the components of the polarization tensors, we select a basis
constructed from the following orthogonal four-vectors $\ $ 
\begin{equation}
h^{\mu }\equiv \frac{\left( \omega ,{\bf k}\right) }{\sqrt{K^{2}}},\text{ \
\ \ \ }l^{\mu }\equiv \frac{\left( k,\omega {\bf n}\right) }{\sqrt{K^{2}}},
\end{equation}
where the space-like unit vector ${\bf n=k}/k$ is directed along the
electromagnetic wave-vector ${\bf k}$. Thus, the longitudinal basis tensor
can be chosen as $L^{\rho \mu }\equiv -l^{\rho }l^{\mu }$ , with
normalization $L_{\rho }^{\rho }=1$. The transverse (with respect to ${\bf k)%
}$ components of $\Pi ^{\,\rho \mu }$ have a tensor structure proportional
to the tensor $T^{\rho \mu }\equiv \left( g^{\rho \mu }-h^{\rho }h^{\mu
}+l^{\rho }l^{\mu }\right) $, where $g^{\rho \mu }={\sf 
\mathop{\rm diag}%
}(1,-1,-1,-1)$ is the signature tensor. This choice of $T^{\rho \mu }$
allows us to describe the two remaining directions orthogonal to $h$ and $l$%
. Therefore, the transverse basis tensor has normalization $T_{\rho }^{\rho
}=2$. One can also check the following orthogonality relations: $l_{\rho
}T^{\rho \mu }=0,$ as well as $k_{i}T^{i\mu }=0,$ and $\ K_{\rho }L^{\rho
\mu }=K_{\rho }T^{\rho \mu }=0$. In this basis, the vector-vector
polarization tensor has the following form 
\begin{equation}
\Pi _{V}^{\,\rho \mu }\left( K\right) =\pi _{l}\left( K\right) L^{\rho \mu
}+\pi _{t}\left( K\right) T^{\rho \mu },  \label{Pi}
\end{equation}
where the longitudinal polarization function is defined as $\pi _{l}\left(
K\right) =\left( 1-\omega ^{2}/k^{2}\right) \Pi _{V}^{\,00}$ and the
transverse polarization function is found to be $\pi _{t}\left( K\right)
=\left( g_{\rho \mu }\Pi _{V}^{\,\rho \mu }-\pi _{l}\right) /2$. The
axial-vector polarizations have to be antisymmetric tensors\footnote{%
We consider also the axial-vector polarization because it will be used in
the next Section.}. They can be written as 
\begin{equation}
\Pi _{AV}^{\,\rho \mu }\left( K\right) =\pi _{AV}\left( K\right) ih_{\lambda
}\epsilon ^{\rho \mu \lambda 0},\ \ \ \ \ \ \ \ \ \ \Pi _{VA}^{\,\rho \mu
}\left( K\right) =\pi _{VA}\left( K\right) ih_{\lambda }\epsilon ^{\rho \mu
\lambda 0},  \label{AV}
\end{equation}
where $\epsilon ^{\rho \mu \lambda 0}$ is the completely antisymmetric
tensor $\left( \epsilon ^{0123}=+1\right) $; $\pi _{AV}\left( K\right) $ and 
$\pi _{VA}\left( K\right) $ are the axial-vector polarization functions of
the medium. As for the axial term, it must be a symmetrical tensor. The most
general expression for this tensor is, therefore 
\begin{equation}
\Pi _{A}^{\mu \nu }\left( K\right) =\pi _{l}\left( K\right) L^{\mu \nu }+\pi
_{t}\left( K\right) T^{\mu \nu }+\pi _{A}\left( K\right) g^{\mu \nu }.
\label{A}
\end{equation}

{\bf Real parts }of the retarded polarization tensors are the same as the
real parts of time-ordered (causal) polarizations, which can be written in
the one-loop approximation as 
\begin{equation}
\Pi ^{\,\mu \rho }=4\pi ie^{2}%
\mathop{\rm Tr}%
\left[ \int \frac{d^{4}p}{(2\pi )^{4}}\,\gamma ^{\mu }\,{\hat{G}}(p)\,\gamma
^{\rho }\,{\hat{G}}(p+K)\right] ,  \label{PT}
\end{equation}
\begin{equation}
\Pi _{VA}^{\,\mu \rho }=4\pi ie^{2}%
\mathop{\rm Tr}%
\left[ \int \frac{d^{4}p}{(2\pi )^{4}}\,\gamma ^{\mu }\,{\hat{G}}(p)\,\gamma
^{\rho }\,\gamma _{5}{\hat{G}}(p+K)\right] ,  \label{PAV}
\end{equation}
\begin{equation}
\Pi _{A}^{\,\mu \rho }=4\pi ie^{2}%
\mathop{\rm Tr}%
\left[ \int \frac{d^{4}p}{(2\pi )^{4}}\,\gamma ^{\mu }\,\gamma _{5}{\hat{G}}%
(p)\,\gamma ^{\rho }\,\gamma _{5}{\hat{G}}(p+K)\right] .  \label{PA}
\end{equation}
Here, ${\hat{G}}(p)$ is the in-medium electron propagator, which includes
the Pauli principle restrictions. This approximation has been studied by
different authors. It corresponds to a collisionless plasma, and thus the
one-loop polarization functions are real-valued in the case $K^{2}>0$, when
Landau damping is not possible. In an ultrarelativistic, strongly-degenerate
electron plasma, the longitudinal and transverse polarization functions take
the form\footnote{%
Our Eq. (\ref{Pil}) differs from Eq. (A39) of the Ref.\cite{BS93} by an
extra factor $\left( \omega ^{2}/k^{2}-1\right) $ because our basis $l^{\mu
} $, $h^{\mu }$ is different from that used by Braaten and Segel. All
components of the complete tensor Eq. (\ref{Pi}) identically coincide with
that obtained in \cite{BS93} for the degenerate case. By the same reason, an
extra factor $\sqrt{K^{2}}$ appears in the $\pi _{VA}$ expression (\ref{PiAV}%
).} \cite{BS93}: 
\begin{equation}
\mathop{\rm Re}%
\pi _{l}=\frac{1}{D_{e}^{2}}\left( 1-\frac{\omega ^{2}}{k^{2}}\right) \left(
1-\frac{\omega }{2k}\ln \frac{\omega +k}{\omega -k}\right) ,  \label{Pil}
\end{equation}
\begin{equation}
\text{\ \ }%
\mathop{\rm Re}%
\pi _{t}=\frac{3}{2}\omega _{pe}^{2}\left[ 1+\left( \frac{\omega ^{2}}{k^{2}}%
-1\right) \left( 1-\frac{\omega }{2kv_{F}}\ln \frac{\omega +k}{\omega -k}%
\right) \right] .  \label{Pit}
\end{equation}
The electron plasma frequency and the Debye screening distance are defined
as 
\begin{equation}
\omega _{pe}^{2}=\frac{4\pi n_{e}e^{2}}{\mu _{e}},\ \ \ \ \ \ \ \ \ \ \ \ \ 
\frac{1}{D_{e}^{2}}=3\omega _{pe}^{2},
\end{equation}
with $\mu _{e}\simeq p_{F}$ and $n_{e}$ being the chemical potential (the
Fermi energy) and \ the number density, respectively, of degenerate
electrons. The axial-vector one-loop polarization tensor is given by 
\begin{equation}
\mathop{\rm Re}%
\pi _{VA}\left( K\right) =\frac{2e^{2}}{\pi }p_{F}\sqrt{K^{2}}\left( 1-\frac{%
\omega ^{2}}{k^{2}}\right) \left( 1-\frac{\omega }{2k}\ln \frac{\omega +k}{%
\omega -k}\right) .  \label{PiAV}
\end{equation}

{\bf Imaginary parts} of polarization tensors appear at next order of
corrections, when electron collisions with ambient particles are included.
The partial contribution of the electron scattering off nuclei to the
imaginary part of the retarded polarizations can be obtained by comparing
the differential rate of the $\nu \bar{\nu}$ production Eq. (\ref{rate11})
with Eq. (\ref{rate2}). We arrive to 
\begin{equation}
\frac{2}{\exp \left( \frac{\omega }{T}\right) -1}%
\mathop{\rm Im}%
\left[ \frac{C_{V}^{2}}{4\pi e^{2}}\Pi _{V}^{\mu \nu }+\frac{C_{A}^{2}}{4\pi
e^{2}}\Pi _{A}^{\mu \nu }\right] =\sum_{i,f}%
\mathop{\rm Tr}%
\left( M^{\dagger \mu }M^{\nu }\right) ,  \label{Opt}
\end{equation}
where summation over initial and final states of the electron should be
understood as described above. This formula can be identified with the
optical theorem. By using the basis expansion Eqs. (\ref{Pi}-\ref{A}) and
the following relations 
\begin{equation}
L_{\mu \nu }L^{\mu \nu }=1,\ \ \ T_{\mu \nu }T^{\mu \nu }=2,\ \ \ L_{\mu \nu
}T^{\mu \nu }=0,\ \ \ L_{\mu \nu }g^{\mu \nu }=1,\ \ \ T_{\mu \nu }g^{\mu
\nu }=2
\end{equation}
we obtain 
\begin{equation}
\mathop{\rm Im}%
\pi _{l}=2\pi e^{2}\frac{\exp \left( \frac{\omega }{T}\right) -1}{\left(
C_{V}^{2}+C_{A}^{2}\right) }\sum_{i,f}\left( L_{\mu \nu }-h_{\mu }h_{\nu
}\right) 
\mathop{\rm Tr}%
\left( M^{\dagger \mu }M^{\nu }\right) ,  \label{l}
\end{equation}
\begin{equation}
\mathop{\rm Im}%
\pi _{t}=\pi e^{2}\frac{\exp \left( \frac{\omega }{T}\right) -1}{\left(
C_{V}^{2}+C_{A}^{2}\right) }\sum_{i,f}\left( T_{\mu \nu }-2h_{\mu }h_{\nu
}\right) 
\mathop{\rm Tr}%
\left( M^{\dagger \mu }M^{\nu }\right) ,  \label{t}
\end{equation}
\begin{equation}
\mathop{\rm Im}%
\pi _{A}=2\pi e^{2}\frac{\exp \left( \frac{\omega }{T}\right) -1}{C_{A}^{2}}%
\sum_{i,f}h_{\mu }h_{\nu }%
\mathop{\rm Tr}%
\left( M^{\dagger \mu }M^{\nu }\right) .  \label{a}
\end{equation}

A direct evaluation of formulae Eq. (\ref{l}-\ref{a}) yield, for the liquid
crust, 
\begin{equation}
\mathop{\rm Im}%
\pi _{l}\left( \omega ,k\right) =-\frac{2}{\pi }\omega _{pe}^{2}Z\alpha ^{2}%
\frac{p_{F}}{k}F_{l}\left( u,\Gamma \right) ,  \label{lPi}
\end{equation}
\begin{equation}
\mathop{\rm Im}%
\pi _{t}\left( \omega ,k\right) =-\frac{1}{\pi }\omega _{pe}^{2}Z\alpha ^{2}%
\frac{p_{F}}{k}F_{0}\left( u,\Gamma \right) -\frac{1}{2}%
\mathop{\rm Im}%
\pi _{l},  \label{tPi}
\end{equation}
\begin{equation}
\mathop{\rm Im}%
\pi _{A}=0.
\end{equation}
The functions $F_{l}\left( u,\Gamma \right) $ and $F_{0}\left( u,\Gamma
\right) $ are defined by the following integrals 
\begin{equation}
F_{l}^{{\sf liquid}}\left( u,\Gamma \right) =\frac{1}{u}\int_{0}^{2p_{F}}%
\frac{dq\,S_{st}\left( q\right) }{\left| \varepsilon \left( q\right) \right|
^{2}q}\left( 1-x^{2}\right) \frac{1}{x^{2}}\left[ 1-\frac{1}{2}\frac{\left(
1-u^{2}\right) }{ux\zeta }\ln \frac{\zeta +xu}{\zeta -xu}\right]
\label{FlLiq}
\end{equation}
\begin{equation}
F_{0}^{{\sf liquid}}\left( u,\Gamma \right) =\int_{0}^{2p_{F}}\frac{%
dq\,S_{st}\left( q\right) }{\left| \varepsilon \left( q\right) \right| ^{2}q}%
\left( 1-x^{2}\right) \left( \frac{1}{x\zeta }\ln \frac{\zeta +xu}{\zeta -xu}%
\right) ,  \label{FtLiq}
\end{equation}
with $x=q/2p_{F}$, and $\zeta =\sqrt{1-u^{2}(1-x^{2})}$. Since we consider
point-like nuclei, the functions $F_{l}\left( u,\Gamma \right) $ and $%
F_{0}\left( u,\Gamma \right) $ depend only on the variable $u=k/\omega $ and
the non-ideality parameter of the ionic component of the plasma \cite{Itoh83}%
: 
\begin{equation}
\Gamma =\frac{Z^{2}e^{2}}{aT}=0.02254\frac{Z^{5/3}}{T_{9}}\frac{p_{F}}{m_{e}}%
,  \label{Gamma}
\end{equation}
where $a$ is the ion-sphere radius{\rm .} Expressions Eq. (\ref{FlLiq}) and
Eq. (\ref{FtLiq}) are valid for $\Gamma <172$. Above this value, the crust
crystallizes. For the crystallin crust we obtain the more complicate
expressions : 
\begin{eqnarray}
F_{l}^{{\sf cryst}}\left( u,\omega ,\Gamma \right) &=&\frac{1}{N_{i}u\omega }%
\int_{0}^{2p_{F}}\frac{dq\,}{\left| \varepsilon \left( q\right) \right| ^{2}q%
}\frac{1-x^{2}}{x^{2}}\left[ 1-\frac{1}{2}\frac{\left( 1-u^{2}\right) }{%
ux\zeta }\ln \frac{\zeta +xu}{\zeta -xu}\right]  \nonumber \\
&&\times \left[ \exp \left( \frac{\omega }{T}\right) -1\right] \int_{-\infty
}^{\infty }dq_{0}\frac{\left( \omega +q_{0}\right) }{\exp \left( \frac{%
\omega +q_{0}}{T}\right) -1}S\left( q_{0},q\right)
\end{eqnarray}
\begin{eqnarray}
F_{0}^{{\sf cryst}}\left( u,\omega ,\Gamma \right) &=&\frac{1}{N_{i}\omega }%
\int_{0}^{2p_{F}}\frac{dq\,}{\left| \varepsilon \left( q\right) \right| ^{2}q%
}\frac{1-x^{2}}{x\zeta }\ln \frac{\zeta +xu}{\zeta -xu}  \nonumber \\
&&\times \left[ \exp \left( \frac{\omega }{T}\right) -1\right] \int_{-\infty
}^{\infty }dq_{0}\frac{\left( \omega +q_{0}\right) }{\exp \left( \frac{%
\omega +q_{0}}{T}\right) -1}S\left( q_{0},q\right) ,
\end{eqnarray}
Inserting the explicit form Eq. (\ref{Sph}) of the one-phonon dynamic form
factor, and performing the integration over $q_{0}$ we obtain 
\begin{eqnarray}
F_{l}^{{\sf cryst}}\left( u,\omega ,\Gamma \right) &=&\frac{\pi }{%
M_{i}u\omega }\left[ \exp \left( \frac{\omega }{T}\right) -1\right] \sum_{%
{\bf s}\lambda {\bf K}}\frac{1}{\omega _{\lambda {\bf s}}}  \nonumber \\
&&\times \int_{0}^{2p_{F}}\frac{dq\,e^{-2W\left( q\right) }\left( {\bf %
q\cdot \hat{e}}_{\lambda {\bf s}}\right) }{\left| \varepsilon \left(
q\right) \right| ^{2}q}\frac{1-x^{2}}{x^{2}}\left( 1-\frac{1-u^{2}}{2ux\zeta 
}\ln \frac{\zeta +xu}{\zeta -xu}\right)  \nonumber \\
&&\times \left[ \frac{\delta \left( {\bf q-s+K}\right) }{1-e^{-\omega
_{\lambda {\bf s}}/T}}\frac{\left( \omega +\omega _{\lambda {\bf s}}\right) 
}{\exp \left( \frac{\omega +\omega _{\lambda {\bf s}}}{T}\right) -1}+\frac{%
\delta \left( {\bf q+s+K}\right) }{e^{-\omega _{\lambda {\bf s}}/T}-1}\frac{%
\left( \omega -\omega _{\lambda {\bf s}}\right) }{\exp \left( \frac{\omega
-\omega _{\lambda {\bf s}}}{T}\right) -1}\right]
\end{eqnarray}
\begin{eqnarray}
F_{0}^{{\sf cryst}}\left( u,\omega ,\Gamma \right) &=&\frac{\pi }{%
M_{i}\omega }\left[ \exp \left( \frac{\omega }{T}\right) -1\right] \sum_{%
{\bf s}\lambda {\bf K}}\frac{1}{\omega _{\lambda {\bf s}}}  \nonumber \\
&&\times \int_{0}^{2p_{F}}\frac{dq\,e^{-2W\left( q\right) }\left( {\bf %
q\cdot \hat{e}}_{\lambda {\bf s}}\right) }{\left| \varepsilon \left(
q\right) \right| ^{2}q}\frac{1-x^{2}}{x\zeta }\ln \frac{\zeta +xu}{\zeta -xu}
\nonumber \\
&&\times \left[ \frac{\delta \left( {\bf q-s+K}\right) }{1-e^{-\omega
_{\lambda {\bf s}}/T}}\frac{\left( \omega +\omega _{\lambda {\bf s}}\right) 
}{\exp \left( \frac{\omega +\omega _{\lambda {\bf s}}}{T}\right) -1}+\frac{%
\delta \left( {\bf q+s+K}\right) }{e^{-\omega _{\lambda {\bf s}}/T}-1}\frac{%
\left( \omega -\omega _{\lambda {\bf s}}\right) }{\exp \left( \frac{\omega
-\omega _{\lambda {\bf s}}}{T}\right) -1}\right] ,
\end{eqnarray}

\section{Including collective effects}

By the Fluctuation-Dissipation Theorem, the function 
\begin{equation}
\Phi ^{\mu \nu }\left( K\right) =\frac{2}{\exp \left( \frac{\omega }{T}%
\right) -1}%
\mathop{\rm Im}%
\Pi ^{\mu \nu }\left( K\right)
\end{equation}
should be identified with the Fourier transform of the correlation function
of two distant weak-currents in the plasma. However, the above
approximation, with $\Pi ^{\mu \nu }$ being the irreducible polarization
tensor, does not take into account for collective effects in the correlation
function. To generalize the correlation function to this case, we should
also include the exchange of an intermediate virtual photon between
electrons. This can be done by summation of the two diagrams shown in Fig.
1, where the thick dashed line is the in-medium photon propagator $D^{\rho
\lambda }\left( K\right) $, defined as the infinite sum of graphs shown in
Fig. 2. In this figure, the thin dashed-line represents ${\rm D}_{\rho
\lambda }^{0}\left( K\right) $ -{\rm \ }the photon propagator in vacuum.{\rm %
\ }According to this series, the exact photon propagator in the medium
satisfies the Dyson's equation 
\begin{equation}
D_{\lambda \rho }\left( K\right) ={\rm D}_{\lambda \rho }^{0}\left( K\right)
+\frac{1}{4\pi }{\rm D}_{\lambda \nu }^{0}\left( K\right) \Pi _{V}^{\,\nu
\mu }\left( K\right) D_{\mu \rho }\left( K\right)
\end{equation}
Within the Lorentz gauge, ${\rm D}_{\lambda \nu }^{0}\left( K\right) $ has
the following form 
\begin{equation}
{\rm D}_{\lambda \rho }^{0}\left( K\right) =\frac{4\pi }{K^{2}}\left(
g_{\lambda \rho }-h_{\lambda }h_{\rho }\right) ,
\end{equation}
The compact block $\Pi _{V}^{\mu \nu }\left( K\right) $ is the retarded
polarization tensor of the plasma, given by Eq. (\ref{Pi}). The retarded
propagator of the in-medium photon has the same tensor structure as the
vector-vector polarization tensor. The solution to the Dyson's equation is
of the form: 
\begin{equation}
D_{\lambda \rho }\left( K\right) =D_{l}\left( K\right) L_{\lambda \rho
}+D_{t}\left( K\right) T_{\lambda \rho }  \label{D}
\end{equation}
with 
\begin{equation}
D_{l}\left( K\right) =\frac{4\pi }{K^{2}-\pi _{l}\left( K\right) },\ \ \ \ \
\ \ \ \ \ \ D_{t}\left( K\right) =\frac{4\pi }{K^{2}-\pi _{t}\left( K\right) 
}.  \label{Dlt}
\end{equation}
Thus, the sum of graphs of the Fig. 1 gives the new tensors : 
\begin{eqnarray}
\tilde{\Pi}_{V}^{\mu \nu }\left( {\bf k},\omega \right) &=&\Pi _{V}^{\mu \nu
}\left( {\bf k},\omega \right) +\frac{1}{4\pi }\Pi _{V}^{\mu \lambda }\left( 
{\bf k},\omega \right) D_{\lambda \rho }\left( {\bf k},\omega \right) \Pi
_{V}^{\rho \nu }\left( {\bf k},\omega \right) ,  \nonumber \\
\tilde{\Pi}_{A}^{\mu \nu }\left( {\bf k},\omega \right) &=&\Pi _{A}^{\mu \nu
}\left( {\bf k},\omega \right) +\frac{1}{4\pi }\Pi _{AV}^{\mu \lambda
}\left( {\bf k},\omega \right) D_{\lambda \rho }\left( {\bf k},\omega
\right) \Pi _{VA}^{\rho \nu }\left( {\bf k},\omega \right) .  \label{DefK}
\end{eqnarray}
By substituting Eqs. (\ref{Pi}-\ref{A}) , (\ref{D}), and (\ref{Dlt}) into
Eqs. (\ref{DefK}) we obtain 
\begin{equation}
\tilde{\Pi}_{V}^{\mu \nu }=\frac{K^{2}\pi _{l}}{K^{2}-\pi _{l}}L^{\mu \nu }+%
\frac{K^{2}\pi _{t}}{K^{2}-\pi _{t}}T^{\mu \nu },  \label{VP}
\end{equation}
\begin{equation}
\tilde{\Pi}_{AV}^{\mu \nu }=\frac{K^{2}}{K^{2}-\pi _{t}}\Pi _{AV}^{\mu \nu
},\ \ \ \ \ \ \ \ \ \ \ \tilde{\Pi}_{VA}^{\mu \nu }=\frac{K^{2}}{K^{2}-\pi
_{t}}\Pi _{VA}^{\mu \nu },  \label{AVP}
\end{equation}
\begin{equation}
\tilde{\Pi}_{A}^{\mu \nu }=\pi _{l}L^{\mu \nu }+\pi _{t}T^{\mu \nu }+\pi
_{A}g^{\mu \nu }-\frac{\pi _{AV}\pi _{VA}}{K^{2}-\pi _{t}}\epsilon ^{\mu
\lambda \sigma 0}\epsilon ^{\rho \nu \tau 0}g_{\lambda \rho }h_{\sigma
}h_{\tau }.  \label{AP}
\end{equation}
Replacement of the irreducible tensors in Eq. (\ref{IrrP}) by the
generalized expressions (\ref{VP}-\ref{AP}) results in the following weak
current-current correlation tensor 
\begin{eqnarray}
\tilde{\Phi}^{\mu \nu }\left( K\right) &=&\frac{2}{\exp \left( \frac{\omega 
}{T}\right) -1}%
\mathop{\rm Im}%
\left[ \frac{C_{V}^{2}}{4\pi e^{2}}\left( \frac{K^{2}\pi _{l}}{K^{2}-\pi _{l}%
}L^{\mu \nu }+\frac{K^{2}\pi _{t}}{K^{2}-\pi _{t}}T^{\mu \nu }\right) \right.
\nonumber \\
&&\left. +\frac{C_{A}^{2}}{4\pi e^{2}}\left( \pi _{l}L^{\mu \nu }+\pi
_{t}T^{\mu \nu }-\frac{\pi _{AV}\pi _{VA}}{K^{2}-\pi _{t}}\epsilon ^{\mu
\lambda \sigma 0}\epsilon ^{\rho \nu \tau 0}g_{\lambda \rho }h_{\sigma
}h_{\tau }\right) \right] .  \label{CF}
\end{eqnarray}

\section{Energy loss due to neutrino-pair radiation}

We consider the emissivity of neutrino pairs from the plasma , i.e. the
total energy which is emitted into neutrino pairs per unit volume and time.
Taking into account collective effects, to the lowest order in $G_{F}$, the
emissivity is given by the following formula: 
\begin{equation}
Q=\frac{G_{F}^{2}}{2}\frac{4\pi }{3}\,\frac{1}{\left( 2\pi \right) ^{6}}%
\sum_{\nu }\int_{0}^{\infty }d\omega \,\omega \int_{k<\omega }d^{3}k\,\left(
K_{\mu }K_{\nu }-K^{2}g_{\mu \nu }\right) \tilde{\Phi}^{\mu \nu }\left(
K\right) .  \label{QTr}
\end{equation}
The symbol $\sum_{\nu }$\ \ indicates that summation over the three neutrino
types has to be performed, with the corresponding values of $C_{V}$\ and $%
C_{A}$, as explained in Section 2. In what follows, this summation will be
understood in any formula of the emissivity. Contraction of the correlation
function (\ref{CF}) with the neutrino-pair tensor $K_{\mu }K_{\nu
}-K^{2}g_{\mu \nu }$, as indicated in Eq. (\ref{QTr}), yields:

\begin{eqnarray*}
&&\left( K_{\mu }K_{\nu }-K^{2}g_{\mu \nu }\right) \tilde{\Phi}^{\mu \nu
}\left( K\right) =-K^{2}\frac{1}{4\pi e^{2}}\frac{2}{\exp \left( \frac{%
\omega }{T}\right) -1}\times \\
&&%
\mathop{\rm Im}%
\left[ C_{V}^{2}K^{2}\left( \frac{\pi _{l}}{K^{2}-\pi _{l}}+2\frac{\pi _{t}}{%
K^{2}-\pi _{t}}\right) +2C_{A}^{2}\frac{k^{2}}{K^{2}}\frac{\pi _{AV}\pi _{VA}%
}{K^{2}-\pi _{t}}+C_{A}^{2}\left( \pi _{l}+2\pi _{t}\right) \right] .
\end{eqnarray*}
Taking the imaginary part of the right-hand side of this equation, after a
lengthy (although straightforward) calculation, we obtain the following
formula : 
\begin{eqnarray}
&&\left( K_{\mu }K_{\nu }-K^{2}g_{\mu \nu }\right) \tilde{\Phi}^{\mu \nu
}\left( K\right) =K^{2}\frac{1}{4\pi e^{2}}\frac{2}{\exp \left( \frac{\omega 
}{T}\right) -1}\times  \nonumber \\
&&\left[ C_{V}^{2}K^{4}\left( \frac{\left| 
\mathop{\rm Im}%
\pi _{l}\right| }{\left( K^{2}-%
\mathop{\rm Re}%
\pi _{l}\right) ^{2}+\left( 
\mathop{\rm Im}%
\pi _{l}\right) ^{2}}+2\frac{\left| 
\mathop{\rm Im}%
\pi _{t}\right| }{\left( K^{2}-%
\mathop{\rm Re}%
\pi _{t}\right) ^{2}+\left( 
\mathop{\rm Im}%
\pi _{t}\right) ^{2}}\right) \right.  \nonumber \\
&&+C_{A}^{2}\left( \left| 
\mathop{\rm Im}%
\pi _{l}\right| +2\left| 
\mathop{\rm Im}%
\pi _{t}\right| \right)  \nonumber \\
&&\left. +2C_{A}^{2}\frac{k^{2}}{K^{2}}\frac{\left| 
\mathop{\rm Im}%
\pi _{t}\right| }{\left( K^{2}-%
\mathop{\rm Re}%
\pi _{t}\right) ^{2}+\left( 
\mathop{\rm Im}%
\pi _{t}\right) ^{2}}\left( 
\mathop{\rm Re}%
\pi _{AV}\right) ^{2}\right] .  \label{contrandim}
\end{eqnarray}
Here it is assumed that $%
\mathop{\rm Im}%
\pi _{A}=%
\mathop{\rm Im}%
\pi _{AV}=0$ , and we incorporated the fact that the imaginary parts of
retarded polarization functions are negative for $\omega >0$, so that $-%
\mathop{\rm Im}%
\pi _{l,t}=\left| 
\mathop{\rm Im}%
\pi _{l,t}\right| $.

As the polarization functions Eqs. (\ref{Pil}, \ref{Pit}) and Eqs. (\ref{lPi}%
, \ref{tPi}) are independent of the direction of the vector ${\bf k}$,
integration over angles can be done in Eq. (\ref{QTr}), and we obtain the $%
\nu \bar{\nu}$ emissivity from the plasma as follows 
\begin{eqnarray}
Q &=&\frac{G_{F}^{2}}{48\pi ^{5}e^{2}}\int_{0}^{\infty }d\omega \,\omega
\int_{0}^{\omega }dk\,\frac{k^{2}K^{2}}{\exp \left( \frac{\omega }{T}\right)
-1}\times  \nonumber \\
&&\left[ C_{V}^{2}K^{4}\left( \frac{\left| 
\mathop{\rm Im}%
\pi _{l}\right| }{\left( K^{2}-%
\mathop{\rm Re}%
\pi _{l}\right) ^{2}+\left( 
\mathop{\rm Im}%
\pi _{l}\right) ^{2}}+2\frac{\left| 
\mathop{\rm Im}%
\pi _{t}\right| }{\left( K^{2}-%
\mathop{\rm Re}%
\pi _{t}\right) ^{2}+\left( 
\mathop{\rm Im}%
\pi _{t}\right) ^{2}}\right) \right.  \nonumber \\
&&+C_{A}^{2}\left( \left| 
\mathop{\rm Im}%
\pi _{l}\right| +2\left| 
\mathop{\rm Im}%
\pi _{t}\right| \right)  \nonumber \\
&&\left. +2C_{A}^{2}\frac{k^{2}}{K^{2}}\left( 
\mathop{\rm Re}%
\pi _{AV}\right) ^{2}\frac{\left| 
\mathop{\rm Im}%
\pi _{t}\right| }{\left( K^{2}-%
\mathop{\rm Re}%
\pi _{t}\right) ^{2}+\left( 
\mathop{\rm Im}%
\pi _{t}\right) ^{2}}\right] .  \label{Qtot}
\end{eqnarray}
This formula describes the $\nu \bar{\nu}$ emissivity from the plasma with
inclusion of collective effects due to the plasma polarization. As we
included the in-medium photon contribution to the correlation function, Eq. (%
\ref{Qtot}) describes the energy losses due to plasmon decay and $\nu \bar{%
\nu}$ Bremsstrahlung from electrons. Due to collective effects, the
contribution to these particular processes through the vector weak-current
can not be separated from one another. This can be understood as follows.
The vector transition current of the electron is responsible for production
of neutrino pairs and on-shell photons in the medium as well. When a virtual
photon participating in the $\nu \bar{\nu}$\ Bremsstrahlung goes on shell,
it has such a large life-time, that decay of the photon into neutrino pairs
can be dually interpreted as free plasmon decay, or as $\nu \bar{\nu}$\
Bremsstrahlung from electrons. This interference does not permit to separate
the partial contributions. However, such an interference can be neglected in
some limiting cases, when one of the processes strongly dominates.

\section{Limiting cases and numerical tests}

\subsection{Moderate temperatures}

At moderate temperatures $T\sim \omega _{pe}$, plasmon decay into
neutrino-pairs strongly dominates the neutrino energy losses from the
degenerate plasma of a liquid crust \cite{IMH92}. In this case, the
imaginary parts of the polarization functions result only in a minor
widening of the plasmon spectra due to plasmon capture and creation in
electron collisions with nuclei. By taking in Eq. (\ref{Qtot}) the imaginary
part of polarization functions equal to zero, we obtain:

\begin{equation}
Q=Q_{l\rightarrow \nu \bar{\nu}}+Q_{t\rightarrow \nu \bar{\nu}},
\end{equation}
where 
\begin{equation}
Q_{l\rightarrow \nu \bar{\nu}}\equiv \frac{G_{F}^{2}C_{V}^{2}}{48\pi
^{4}e^{2}}\int_{0}^{\infty }\frac{d\omega \,\omega }{\exp \left( \frac{%
\omega }{T}\right) -1}\int_{0}^{\omega }dk\,k^{2}K^{6}\delta \left(
K^{2}-\pi _{l}\right) ,  \label{Ql}
\end{equation}
\begin{equation}
Q_{t\rightarrow \nu \bar{\nu}}\equiv \frac{G_{F}^{2}}{24\pi ^{4}e^{2}}%
\int_{0}^{\infty }\frac{d\omega \,\omega }{\exp \left( \frac{\omega }{T}%
\right) -1}\int_{0}^{\omega }dk\,k^{2}\left(
C_{V}^{2}K^{6}+C_{A}^{2}k^{2}\pi _{AV}^{2}\right) \delta \left( K^{2}-\pi
_{t}\right) .  \label{Qt}
\end{equation}
As it follows from the poles of Eq. (\ref{Dlt}), longitudinal and transverse
eigen-photon modes satisfy the dispersion relations 
\begin{equation}
\omega ^{2}-k^{2}-%
\mathop{\rm Re}%
\pi _{l,t}(\omega ,k)=0,  \label{Disp}
\end{equation}
therefore Eqs. (\ref{Ql}) and (\ref{Qt}) are easily interpreted as the
emissivity due to decay of on-shell longitudinal $\left( l\right) $ and
transverse $\left( t\right) $ plasmons, respectively, into neutrino pairs.
We will now show that, in fact, Eqs. (\ref{Ql}) and (\ref{Qt}) give the
known results for the neutrino pair emissivity from longitudinal and
transverse plasmons, respectively.

By expanding $K^{2}-%
\mathop{\rm Re}%
\pi _{t}$ around $\omega _{t}$ one can approximate

\begin{equation}
\omega ^{2}-k^{2}-%
\mathop{\rm Re}%
\pi _{l,t}\left( \omega ,k\right) \simeq \frac{1}{Z_{l,t}}\left( \omega
^{2}-\omega _{l,t}^{2}\left( k\right) \right)
\end{equation}
with 
\begin{equation}
Z_{l,t}(k)\equiv \left( 1-\left. \frac{\partial \pi _{l,t}(\omega ,k)}{%
\partial \omega ^{2}}\right| _{\omega ^{2}=\omega _{l,t}^{2}}\right) ^{-1}.
\end{equation}
In this way 
\begin{equation}
\delta \left( \omega ^{2}-k^{2}-%
\mathop{\rm Re}%
\pi _{l,t}\right) =\frac{Z_{l,t}(k)}{2\omega _{l,t}(k)}\delta \left( \omega
-\omega _{l,t}(k)\right) ,
\end{equation}
where $\omega _{l,t}(k)$ is the solution (for $\omega >0$) of the dispersion
equation (\ref{Disp}) for longitudinal or transverse photons. Substitution
into Eqs. (\ref{Ql}) and (\ref{Qt}) results in 
\begin{eqnarray}
Q_{l\rightarrow \nu \bar{\nu}} &=&C_{V}^{2}\frac{G_{F}^{2}}{96\pi ^{4}e^{2}}%
\int_{0}^{\infty }d\omega \frac{1}{\exp \left( \frac{\omega }{T}\right) -1}%
\int_{0}^{\omega }dk\,k^{2}\,Z_{l}(k)K^{6}\delta \left( \omega -\omega
_{l}(k)\right)  \nonumber \\
Q_{t\rightarrow \nu \bar{\nu}} &=&\frac{G_{F}^{2}}{48\pi ^{4}e^{2}}%
\int_{0}^{\infty }d\omega \int_{0}^{\omega }dk\,k^{2}\frac{Z_{t}(k)}{\exp
\left( \frac{\omega }{T}\right) -1}\left[ C_{V}^{2}K^{6}+C_{A}^{2}k^{2}%
\mathop{\rm Re}%
^{2}\pi _{AV}(\omega ,k)\right] \delta \left( \omega -\omega _{t}(k)\right) .
\end{eqnarray}
The integral over $\omega $ is trivial. As for the integral over $k$, one
has to remember that the condition $K^{2}>0$ implies that, for longitudinal
modes, $k$ can not be larger than the limiting value $k_{\max }$ \cite{BS93}%
. Finally, we obtain 
\begin{eqnarray}
Q_{l\rightarrow \nu \bar{\nu}} &=&\frac{G_{F}^{2}C_{V}^{2}}{96\pi ^{4}e^{2}}%
\int_{0}^{k_{\max }}dk\,\frac{\,k^{2}Z_{l}(k)}{\exp \left( \frac{\omega
_{l}\left( k\right) }{T}\right) -1}\left( \omega _{l}^{2}\left( k\right)
-k^{2}\right) ^{3},  \nonumber \\
Q_{t\rightarrow \nu \bar{\nu}} &=&\frac{G_{F}^{2}}{48\pi ^{4}e^{2}}%
\int_{0}^{\infty }\,dk\frac{\,k^{2}Z_{t}(k)}{\exp \left( \frac{\omega
_{t}\left( k\right) }{T}\right) -1}\left[ C_{V}^{2}\left( \omega
_{t}^{2}\left( k\right) -k^{2}\right) ^{3}+C_{A}^{2}k^{2}%
\mathop{\rm Re}%
\pi _{AV}^{2}\left( \omega _{t}\left( k\right) ,k\right) \right] .
\label{Qtfin}
\end{eqnarray}
These formulae coincide with the known expressions \cite{BS93}\ for neutrino
pair emission from plasmons.

\subsection{High-temperature limit}

We now consider a degenerate electron gas in the case of high temperatures $%
T\gg \omega _{pe}$. Then the emissivity of Bremsstrahlung, which increases
as $T^{6}$, strongly overcomes the emissivity due to plasmon decay, which
only goes as $T^{3}$ (see , e.g. \cite{BS93}). At such temperatures, the
dominant contribution comes from $K^{2}\gg 
\mathop{\rm Re}%
\pi _{l,t}\sim \omega _{pe}^{2}$. Thus, neglecting $%
\mathop{\rm Re}%
\pi _{l,t}$ and $%
\mathop{\rm Im}%
\pi _{l,t}$ in the denominators of Eq. (\ref{Qtot}), and considering also $%
\left( 
\mathop{\rm Re}%
\pi _{AV}\right) ^{2}\ll K^{4}$, we obtain 
\begin{equation}
Q_{Br}=\frac{G_{F}^{2}\left( C_{V}^{2}+C_{A}^{2}\right) }{48\pi ^{5}e^{2}}%
\int_{0}^{\infty }\frac{d\omega \,\omega }{\exp \left( \frac{\omega }{T}%
\right) -1}\int_{0}^{\omega }dk\,k^{2}K^{2}\left( \left| 
\mathop{\rm Im}%
\pi _{l}\right| +2\left| 
\mathop{\rm Im}%
\pi _{t}\right| \right) .
\end{equation}
Performing this integral, with $%
\mathop{\rm Im}%
\pi _{l,t}$ defined by Eqs. (\ref{lPi}, \ref{tPi}), we arrive to the known
result \cite{Itoh83} for the $\nu \bar{\nu}$ Bremsstrahlung emissivity: 
\begin{eqnarray}
Q_{Br} &=&\frac{4\pi }{189}G_{F}^{2}\left( C_{V}^{2}+C_{A}^{2}\right)
e^{4}N_{i}Z^{2}T^{6}  \nonumber \\
&&\times \int_{0}^{2p_{F}}\frac{dq\,S\left( q\right) }{\left| \varepsilon
\left( q\right) \right| ^{2}q}\left( 1-x^{2}\right) \int_{0}^{1}\,du\frac{%
u\left( 1-u^{2}\right) }{x\zeta }\ln \frac{\zeta +xu}{\zeta -xu}.
\end{eqnarray}
The latter integral can be done analytically (see also \cite{HKY96}) : 
\begin{equation}
\int_{0}^{1}du\frac{u\left( 1-u^{2}\right) }{x\zeta }\ln \frac{\zeta +xu}{%
\zeta -xu}=\frac{2}{3}\frac{1}{1-x^{2}}\left( 1+\frac{2x^{2}}{1-x^{2}}\ln
x\right) ,
\end{equation}
and we finally obtain 
\begin{equation}
Q_{Br}=\frac{8\pi }{567}G_{F}^{2}\left( C_{V}^{2}+C_{A}^{2}\right)
e^{4}N_{i}Z^{2}T^{6}\int_{0}^{2p_{F}}\frac{dq\,S\left( q\right) }{\left|
\varepsilon \left( q\right) \right| ^{2}q}\left( 1+\frac{2x^{2}}{1-x^{2}}\ln
x\right) .  \label{HTBr}
\end{equation}

\subsection{Low-temperature limit: collective effects}

In the case of low temperatures $T\ll \omega _{pe}$, the occupation numbers
of eigen-photon modes in the medium are exponentially suppressed, and the
typical energy of a neutrino-antineutrino pair is of the order of the medium
temperature. Under this situation we can neglect $K^{2}\sim T^{2}$ with
respect to $%
\mathop{\rm Re}%
\pi _{l,t}\sim \omega _{pe}^{2}$ in the denominators of Eq. (\ref{Qtot}) .

Now, Eq. (\ref{Qtot}) takes the form 
\begin{equation}
Q=Q^{V}+Q^{A}.
\end{equation}
These terms describe the vector ($Q^{V}$) and an axial ($Q^{A}$)
weak-current contributions 
\begin{equation}
Q^{V}\equiv \frac{G_{F}^{2}C_{V}^{2}}{48\pi ^{5}e^{2}}\int_{0}^{\infty }%
\frac{d\omega \,\omega }{\exp \left( \frac{\omega }{T}\right) -1}%
\int_{0}^{\omega }dk\,k^{2}K^{2}\left( \frac{K^{4}}{%
\mathop{\rm Re}%
^{2}\pi _{l}}\left| 
\mathop{\rm Im}%
\pi _{l}\right| +2\frac{K^{4}}{%
\mathop{\rm Re}%
^{2}\pi _{t}}\left| 
\mathop{\rm Im}%
\pi _{t}\right| \right) ,  \label{QVbr}
\end{equation}
\begin{equation}
Q^{A}\equiv \frac{G_{F}^{2}C_{A}^{2}}{48\pi ^{5}e^{2}}\int_{0}^{\infty }%
\frac{d\omega \,\omega }{\exp \left( \frac{\omega }{T}\right) -1}%
\int_{0}^{\omega }dk\,k^{2}K^{2}\left( \left| 
\mathop{\rm Im}%
\pi _{l}\right| +2\left| 
\mathop{\rm Im}%
\pi _{t}\right| \right) .  \label{QAbr}
\end{equation}
In the latter equation we neglected terms which are proportional to $%
\mathop{\rm Re}%
\pi _{AV}/%
\mathop{\rm Re}%
\pi _{t}\sim T/\mu _{e}\ll 1$.

As mentioned above, the imaginary parts of polarization tensors are due to
electron collisions with nuclei. Thus, Eqs. (\ref{QVbr}, \ref{QAbr}) should
be understood as the neutrino-pair Bremsstrahlung from electrons.
Substituting in Eq. (\ref{QVbr}) the explicit form of the polarizations, and
performing integrations over $dk$ and $d\omega $ we obtain, for a liquid
crust, the following expression for the axial weak-current contribution 
\begin{equation}
Q_{Br}^{A}=\frac{8\pi }{567}G_{F}^{2}C_{A}^{2}e^{4}N_{i}Z^{2}T^{6}%
\int_{0}^{2p_{F}}\frac{dq\,S\left( q\right) }{\left| \varepsilon \left(
q\right) \right| ^{2}q}\left( 1+\frac{2x^{2}}{1-x^{2}}\ln x\right)
\end{equation}
which coincides with the known result Eq. (\ref{HTBr}) for the axial term%
{\rm \ }in the ultrarelativistic case, while the contribution of the vector
weak current contains in the integrand the additional, small factors, $K^{4}/%
\mathop{\rm Re}%
^{2}\pi _{l,t}\sim T^{4}/\omega _{pe}^{4}$ and therefore decreases along
with the temperature as $T^{10}$. Thus, at low temperatures, the
contribution of the vector weak current to emissivity is suppressed due to
collective effects, as discussed in \cite{L99}.

In Fig. 3, we show the result of a numerical calculation of the
neutrino-pair emissivity from a liquid crust with a density $\rho $ such
that $\rho Y_{e}=7.3\times 10^{9}gr\;cm^{-3}$ ($Y_{e}$ is the electron
fraction per baryon), consisting on nuclei with an atomic number $Z=30$
embedded into a degenerate ultrarelativistic electron gas, which is in good
agrement with the total neutrino emissivity, as given by the sum of the
plasmon decay contribution, calculated in \cite{IMH92}, and the ''standard''
Bremsstrahlung contribution calculated in \cite{Itoh83}. One readily
observes from this figure, that our calculation coincides with the standard
Bremsstrahlung emissivity, when this is the dominant process (at high
temperatures), and with plasmon decay at lower temperatures.

Collective effects are dramatic at low temperatures $T\ll \omega _{pe}$,
when plasmon decay is negligible, which are typical for the crystallin
crust. However, the above criterion $\Gamma >172$ for crystallization of the
crust corresponds to the case where an infinite time scale is allowed to
attain thermal equilibrium. In actual stellar evolution, only a finite time
scale is allowed for temperature variation. Therefore, the ionic system is
likely in an amorphous-crystal state for $172<\Gamma \lesssim 210$ \cite
{IIM83}.\ Thus, in practice, the liquid state of the crust persists up to $%
\Gamma \lesssim 210$. At such temperatures, the plasmon decay is
exponentially suppressed, and Bremsstrahlung becomes again dominating. In
this case, collective effects dramatically suppress the vector weak-current
contribution to neutrino-pair Bremsstrahlung from electrons in the liquid
crust.

Unfortunately, the structure factor is not tabulated for the liquid ionic
system in this domain of \ $\Gamma $. For this reason, in order to make
these effects more apparent, we considered a liquid crust with the same
value of $\rho Y_{e}$, consisting on nuclei with $Z=10$. Such a crust
remains liquid for temperatures much smaller than the plasma frequency{\rm .}
The result of our calculation is plotted in Fig. 4, where we show separately
the $Q^{V}$ and $Q^{A}$ pieces, in comparison with plasmon decay. The $Q^{A}$
contribution to the emissivity shows the standard behavior $\sim T^{6}$,
while the $Q^{V}$ piece drops much faster with temperature, as $T^{10}$. As
we discussed above, such a behavior is a consequence of collective effects,
which result on the screening of the vector weak-current of electrons in the
plasma.

In Fig. 5, we have plotted the ratio $R$ of the total emissivity, as
calculated from our formulae, to the sum of the standard Bremsstrahlung plus
plasmon decay (taken from \cite{Itoh83} and \cite{IMH92}). As it readily
seen from this figure, this ratio abruptly decreases at low temperatures,
and tends to the expected value \cite{L99}. 
\begin{equation}
\frac{Q}{Q_{0}}=\frac{C_{A}^{2}+2C_{A}^{\prime 2}}{C_{V}^{2}+C_{A}^{2}+2%
\left( C_{V}^{\prime 2}+C_{A}^{\prime 2}\right) }=0.\,\allowbreak 448.
\label{wRatio}
\end{equation}

\section{Conclusions}

We have shown a method of calculation for the neutrino-pair energy losses
from the crust of a neutron star, which has the advantage that collective
effects are automatically included, and do not need a separate
consideration. The formula we obtained for the neutrino-pair emissivity, Eq.
(\ref{Qtot}), includes the contribution of plasmon decay into neutrino
pairs, as well as neutrino-pair emission due to electron collisions with
nuclei and phonons. Some limiting cases, considered in Sect. 6, demonstrate
that this formula describes the known results for energy losses due to
plasmon decay and neutrino-pair Bremsstrahlung of electrons, when one of
these processes dominates, at moderate or high temperatures. Our formula
takes into account collective effects, which in the case of low temperatures
manifest in a suppression of the vector weak-current contribution to the
emissivity, in accordance with the results in \cite{L99}. Some numerical
tests for moderate and high temperatures show a good agreement of our
calculated total emissivity, with that obtained by summation of the separate
contributions of plasmon decay and electron Bremsstrahlung obtained by
different authors. In general, Eq. (\ref{Qtot}) can also be used for a
non-degenerate plasma, if the corresponding polarization functions are
inserted. Thus, Eq. (\ref{Qtot}) is able to incorporate neutrino-pair energy
losses from the electron plasma caused by processes other than plasmon decay
and neutrino-pair Bremsstrahlung. This can be done by improvement of the
imaginary parts of the retarded polarization functions,{\rm \ }with
inclusion of different physical processes leading to damping of the plasmon
in the medium under consideration.

\acknowledgments
This work has been supported by the Spanish Grants DGES PB97- 1432 and
AEN99-0692. L. B. Leinson would like to thank the Russian Foundation for
Fundamental Research Grant 97-02-16501.

\mafigura{16cm}{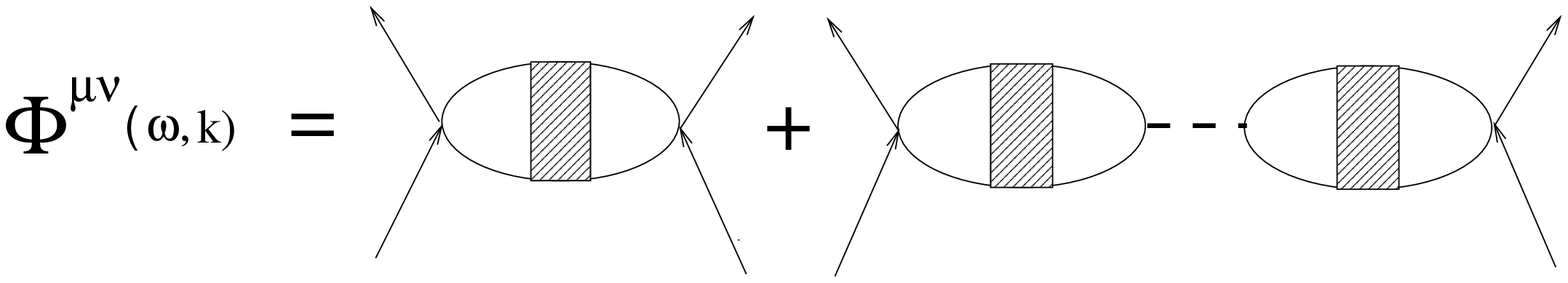}{Diagrams contributing to the weak correlation function, 
when collective effects are included. Shadow rectangles on electron loops 
represent intermediate electron interactions with nuclei (in the liquid
crust) or with phonons (in the crystall). The thick dashed line on the second 
diagram stands for an in-medium photon (see next figure).}{Fig. 1}

\vspace{3cm}

\mafigura{16cm}{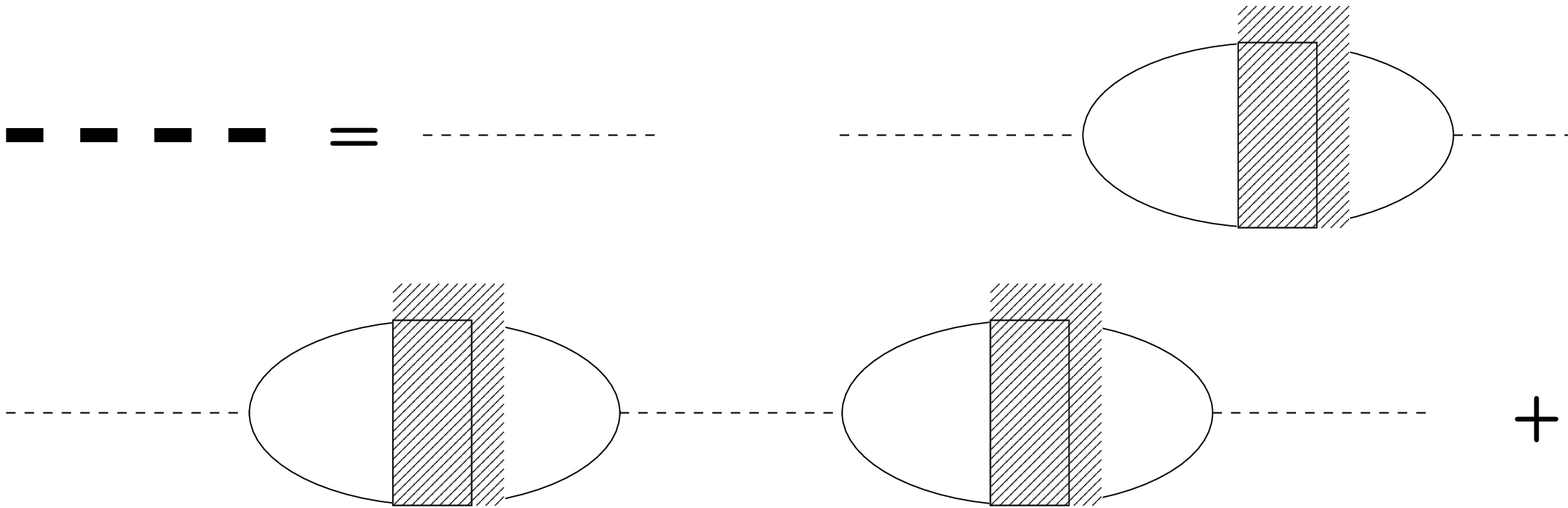}{Series of diagrams defining the in-medium photon propagator (shown as a thick 
dashed line). Thin dashed lines correspond to the in-vacuum photon
propagator.}{Fig. 2}

\newpage

\begin{figure}
\includegraphics[width=12cm,angle=270]{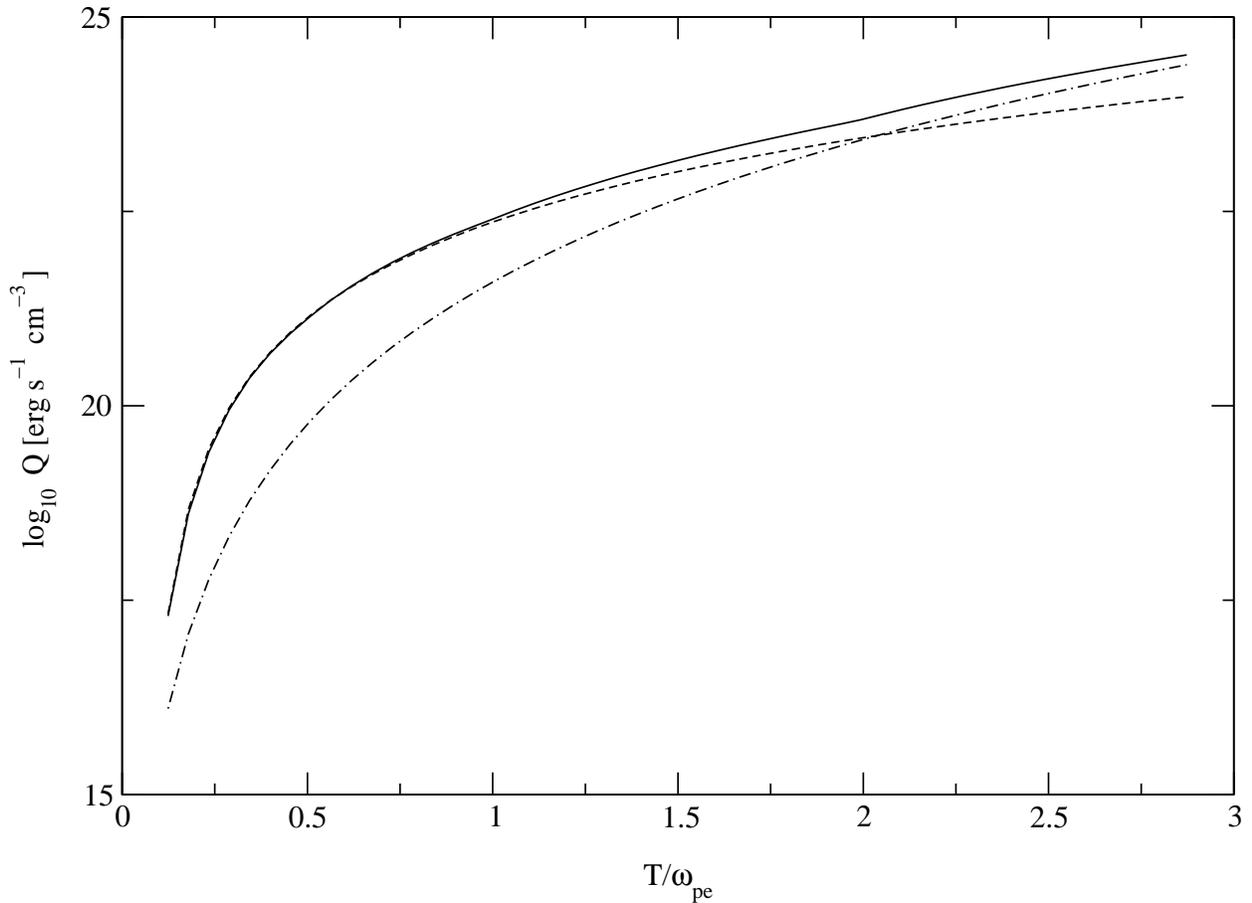}
\caption{Total neutrino-pair emissivity, calculated from our 
Eq. \ref{Qtot}  (solid line), in comparison with "standard" (i.e., without
collective  effects) plasmon decay (dashed line) and Bremsstrahlung
(dashed-dotted line).  All curves are for a density corresponding to $\rho
Y_e=7.3 \times 10^9 gr cm^{-3}$ and an atomic number $Z=30$ for ions.}
\label{Fig. 3}
\end{figure}

\vspace{2cm}

\begin{figure}
\includegraphics[width=12cm,angle=270]{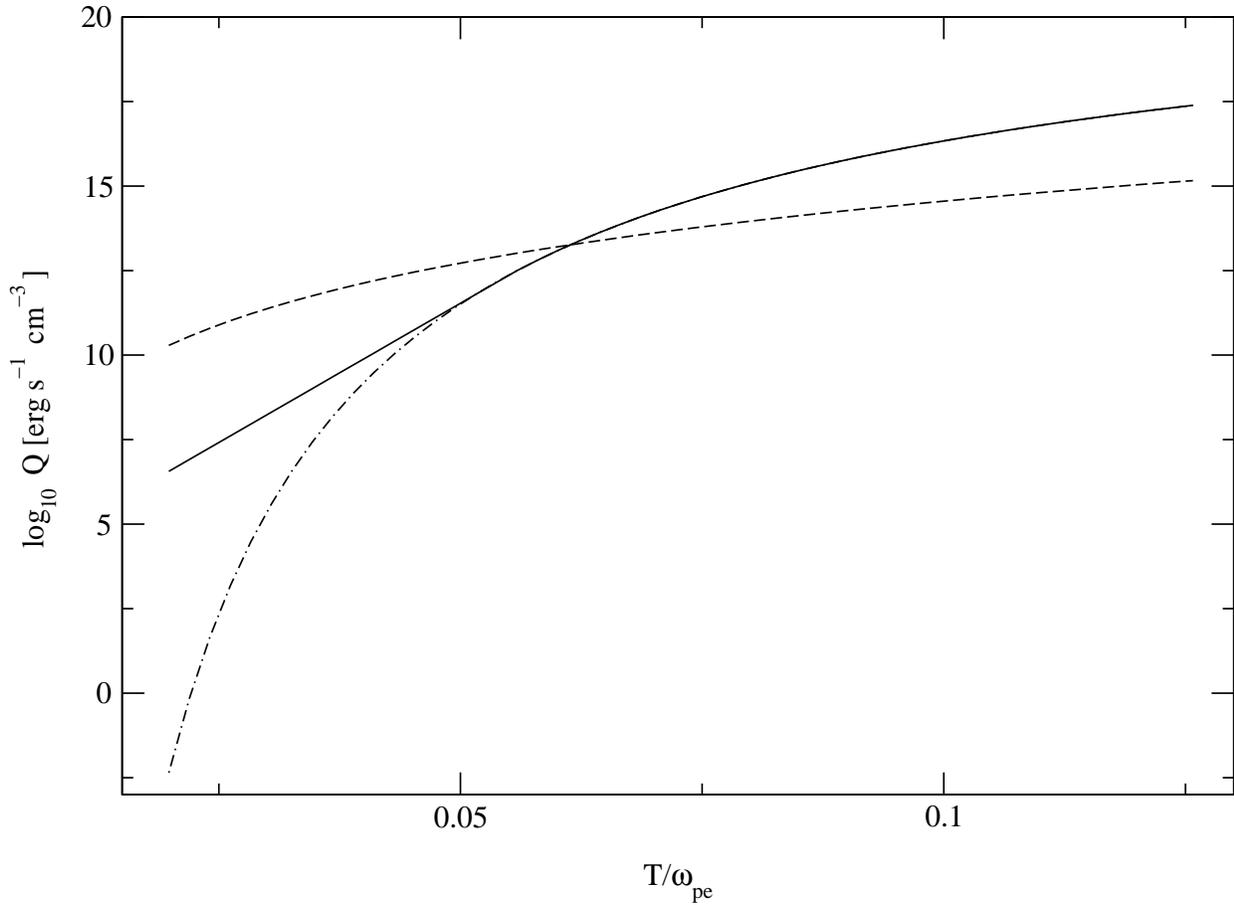}
\caption{Separate vector (solid line) and axial (dashed line) 
contributions to the neutrino-pair emissivity. At low temperatures, the 
vector weak-current contribution is suppressed with respect to the axial 
term due to collective effects. At moderate temperatures, when  
plasmon decay dominates, the vector weak-current contribution coincides with
the plasmon decay curve (dashed-dotted line). All curves correspond the same 
value of $\rho Y_e$ as in the previous figure, and an atomic number $Z=10$ 
for ions.}
\label{Fig. 4}
\end{figure}

\begin{figure}
\includegraphics[width=12cm,angle=270]{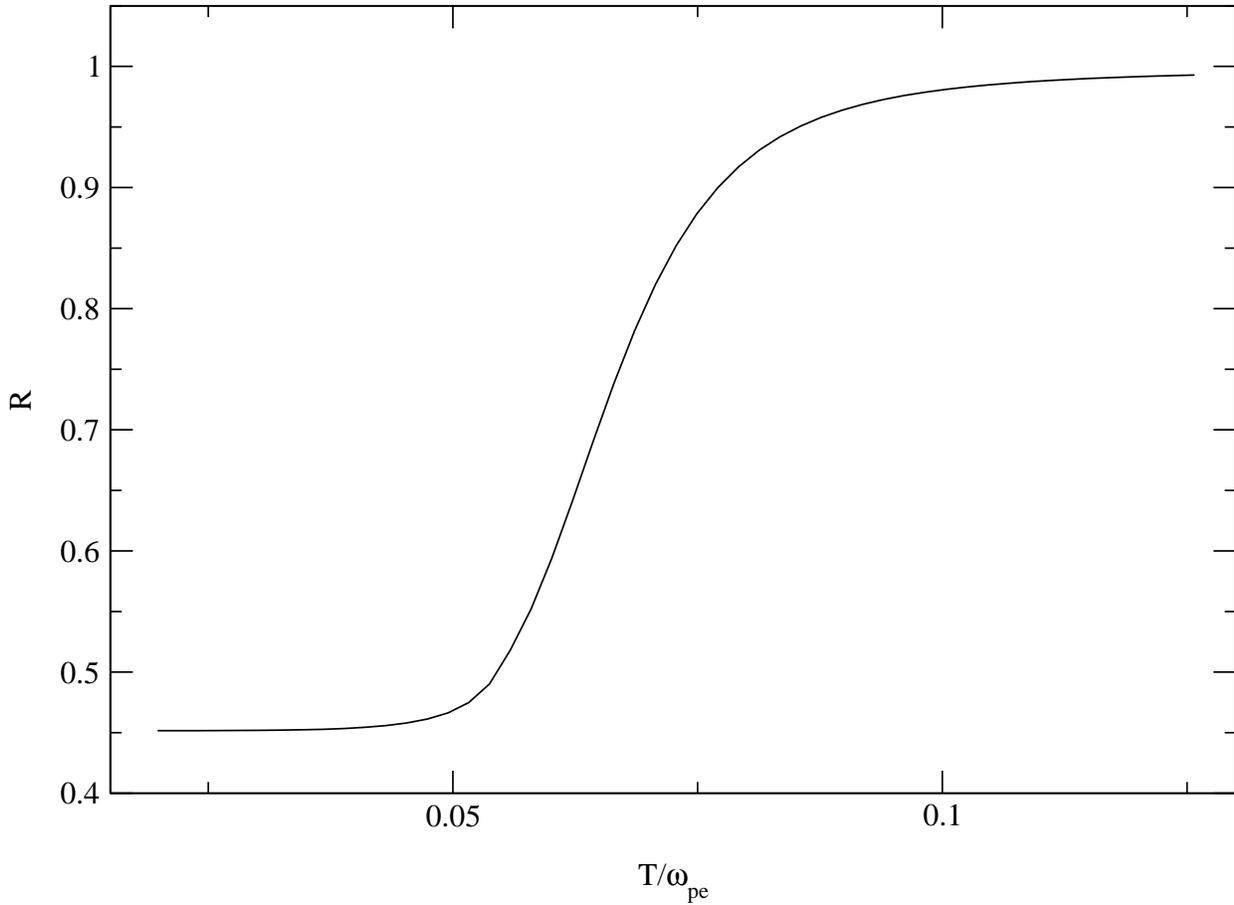}
\caption{The ratio of the total neutrino-pair emissivity, as obtained from our 
formula, to the sum of standard plasmon decay and Bremsstrahlung. 
Plasma conditions are the same as in the previous figure.}
\label{Fig. 5}
\end{figure}

\end{document}